\documentclass[reprint,
amsmath,amssymb,prl
]{revtex4-1}

\usepackage{dcolumn}
\usepackage{bm}
\usepackage{hyperref}

\usepackage[dvipsnames]{xcolor}
\usepackage{tikz}
\usetikzlibrary{calc,matrix,decorations.pathmorphing,decorations.markings,arrows,positioning,intersections,mindmap,backgrounds}
\usepackage[yyyymmdd,hhmmss]{datetime}

\newcommand{\bfn}[2]{{\mathrm{B}\!\left(\begin{matrix}{#1}\\{#2}\end{matrix}\right)}}
\newcommand{\residue}[1]{{\underset{#1}{\text{Res}}}}

\begin{document}


\title{Loops in the Bulk}

\author{Ellis Ye Yuan}%
\email{yyuan@ias.edu}
\affiliation{%
School of Natural Sciences, Institute for Advanced Study, Princeton, NJ 08540, USA
}%
\date{\today}

\begin{abstract}
We initiate a systematic investigation of Mellin amplitudes of Witten diagrams to all loop levels, by introducing integral recursion relations among them. Focusing on the scalar effective theories in AdS with the simplest type of interactions, the integral kernel that triggers the recursion obeys universal rules. As a first application, analytic properties of a 4-point triangle diagram are analyzed with this method.
\end{abstract}

\maketitle


\section{\label{sec:introduction}Introduction}

Perturbative study of scattering in AdS using Witten diagrams is interesting not only because this is the simplest playground probing dynamics of weakly coupled QFTs in curved spacetime, but also due to the need in precision checks of AdS/CFT correspondence \cite{Maldacena:1997re,Gubser:1998bc,Witten:1998qj} (for a useful review on this aspect, e.g., \cite{DHoker:2002nbb}). As the AdS bulk isometries induce conformal symmetries on its boundary, Witten diagrams enjoy simpler and useful representation when transformed from (boundary) spacetime to Mellin space, which is analogous to the momentum space for Feynman diagrams \cite{Mack:2009mi,Mack:2009gy}. Specifically, focusing on scalar operators, any boundary correlator can be written into the form
\begin{equation}\label{eq:MellinDef}
\langle\mathcal{O}_1(x_1)\cdots\mathcal{O}_n(x_n)\rangle=\!\!\int[\mathrm{d}\delta]\mathcal{M}(\delta)\prod_{i<j}^n\frac{\Gamma[\delta_{i\,j}]}{(x_i-x_j)^{2\delta_{i\,j}}},
\end{equation}
which defines its associated \emph{Mellin amplitude} $\mathcal{M}(\delta)$. The $\delta$'s are the Mellin space variables, satisfying
\begin{equation}
\delta_{i\,j}=\delta_{j\,i},\quad
\delta_{i\,i}=-\Delta_i,\quad
\sum_{j=1}^n\delta_{i\,j}=0,\quad
1\leq i\leq n,
\end{equation}
where $\Delta_i$ denotes the dimension of $\mathcal{O}_i$. The measure $[\mathrm{d}\delta]$ in \eqref{eq:MellinDef} is the product of $\frac{\mathrm{d}\delta}{2\pi i}$ over any choice of independent set, and each $\delta$ contour stretches between $\pm i\infty$, separating the left and right poles (depending on the signs in front of the variable) of the integrand. Mellin amplitude has the advantage of being meromorphic as long as the correlator observes discrete spectrum in any OPE channel, with its poles corresponding to operators appearing in the OPE. Even better, in the existence of a large $N$ expansion (as required by holography \cite{Heemskerk:2009pn}) multi-trace operators formed by $\mathcal{O}$'s are clearly distinguished from all other operators in the OPE, as the former are captured by the additional product in RHS of \eqref{eq:MellinDef}. This allows $\mathcal{M}$ to cleanly encode dynamics inside the bulk, i.e., operators $\underline{\mathcal{O}}$ essentially tied to the bulk propagating fields \cite{Penedones:2010ue,Fitzpatrick:2011ia}. In fact, $\mathcal{M}$ depends on the Mellin variables only via their Mandelstam-like combinations (we will simply call them Mandelstam variables), $\mathcal{M}\equiv\mathcal{M}(s)$, with
\begin{equation}
s_A=\sum_{i\in A}\Delta_i-2\sum_{i<j\in A}\delta_{i\,j},
\end{equation}
where $A$ denotes a set of boundary points. Each $s_A$ associates to an OPE channel that separates the points in $A$ and those in its complement $A^{\rm c}$. For later convenience we refer \emph{non-trivial (OPE) channel} to that whose Mandelstam variable enters $\mathcal{M}$.

While at tree level Mellin amplitudes can be easily worked out by the well-established diagrammatic rules \cite{Fitzpatrick:2011ia,Paulos:2011ie,Fitzpatrick:2011hu,Nandan:2011wc}, beyond that available results are limited to only several simplest cases \cite{Penedones:2010ue,Aharony:2016dwx}, essentially due to the difficulty of bulk integrations in the straightforward computation (see, e.g., \cite{Giombi:2017hpr,Cardona:2017tsw} for more recent discussions).

\begin{figure}[ht]
\begin{center}
\begin{tikzpicture}
\begin{scope}[scale=.7]
\begin{scope}[xshift=-3cm]
\draw [black,thick] (0,0) -- (0:2);
\draw [black,thick] (0,0) -- (60:2);
\draw [black,thick] (0,0) -- (120:2);
\draw [black,thick] (0,0) -- (180:2);
\draw [black,very thick,dotted] (-90:2) .. controls (-1,-1.5) and (-1,-.5) .. (0,0) .. controls (1,-.5) and (1,-1.5) .. (-90:2);
\draw [black,fill=black!15!white] (0,0) circle [radius=.85];
\draw [black,ultra thick] (0,0) circle [radius=2];
\fill [black] (-90:2) circle [radius=3pt];
\node [anchor=north] at (-90:2) {\scriptsize $x$};
\end{scope}
\node [anchor=center] at (0,0) {\Large $\longrightarrow$};
\node [anchor=south] at (0,.2) {\scriptsize $\displaystyle\int\frac{[\mathrm{d}c]_{\underline\Delta}}{\mathcal{C}_{h\pm c}}$};
\node [anchor=north] at (0,-.1) {\scriptsize $\displaystyle\int\mathrm{d}^dx$};
\begin{scope}[xshift=3cm]
\draw [black,thick] (0,0) -- (0:2);
\draw [black,thick] (0,0) -- (60:2);
\draw [black,thick] (0,0) -- (120:2);
\draw [black,thick] (0,0) -- (180:2);
\draw [black,very thick,dotted] (0,0) .. controls (1,-.5) and (.8,-1.5) .. (-90:1.5) .. controls (-.8,-1.5) and (-1,-.5) .. (0,0);
\node [anchor=south] at (0,-1.5) {\scriptsize $\underline\Delta$};
\draw [black,fill=black!15!white] (0,0) circle [radius=.85];
\draw [black,ultra thick] (0,0) circle [radius=2];
\end{scope}
\end{scope}
\end{tikzpicture}
\end{center}
\vspace{-1.5em}\caption{\label{fig:loopformation}Loop formation.}
\end{figure}
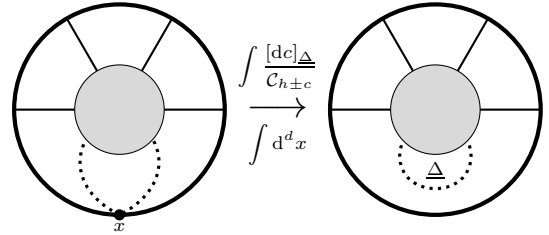

Nevertheless, given the convenience of Mellin space, it is best to make full use of its power and carry out computations solely in it, for the study of loop diagrams. The basic idea is to introduce recursion relations that brings a given Mellin amplitude to a new one at one higher loop, as depicted in Figure \ref{fig:loopformation}.
While pictorially this way of forming a loop resembles taking forward limit for scattering in Minkowski \cite{Feynman:1963ax}, in AdS this is fulfilled by the split representation \cite{Penedones:2010ue}, which (in the scalar case) expresses a given bulk-to-bulk propagator of dimension $\underline\Delta$ as a boundary integral and a spectrum integral
\begin{equation}\label{eq:splitrepresentation}
G_{\rm bb}^{\underline\Delta}(X,\!Y)\!=\!\!\!\int\!\frac{[\mathrm{d}c]_{\underline\Delta}\mathcal{N}_c}{\mathcal{C}_{h\pm c}}\!\!\!\!\underset{\partial\text{AdS}}{\int}\!\!\!\mathrm{d}^dx\,G_{\rm{b}\partial}^{h-c}(X,x)G_{\rm{b}\partial}^{h+c}(Y,x),
\end{equation}
with $\mathcal{C}_\Delta$ being the normalization for $G_{\text{b}\partial}^{\Delta}$, and
\begin{equation}
[\mathrm{d}c]_{\underline\Delta}=\frac{\mathrm{d}c}{2\pi i}\frac{1}{(\underline\Delta-h)^2-c^2},\quad
\mathcal{N}_c=\frac{\Gamma(h\pm c)}{2\pi^{2h}\Gamma(\pm c)},
\end{equation}
where we abbreviate $f(\pm c)\equiv f(+c)f(-c)$ for any notation $f$. The contour for the spectrum variable $c$ is define to be left to the pole at $c=\underline\Delta-h$ and right to $c=h-\underline\Delta$, again stretched between $\pm i\infty$.
 
Assuming the complete knowledge about an existing diagram, we can identify the conformal dimension of two of its boundary points as $h\pm c$ respectively, and take their coincidence limit at point $x$. After integrating $x$ over the entire boundary and further carrying out a spectrum integration of $c$ according to \eqref{eq:splitrepresentation} with some value $\underline\Delta$, we obtain a new diagram where the two bulk-to-boundary propagators of the original diagram glue into a bulk-to-bulk propagator of dimension $\underline\Delta$. While the loop formation as stated above is an operation in spacetime, it can be translated into a well-defined operation on the Mellin amplitude, schematically of the form
\begin{equation}\label{eq:recursionschematic}
\mathcal{M}(\delta)=\int[\mathrm{d}\tau]\,\mathcal{M}_0(\tau)\,\mathcal{K}(\tau,\delta).
\end{equation}
Attention that in switching from the original diagram to the new diagram we also move into a completely different Mellin space. To distinguish, we use  $\delta$ for Mellin variables of the new diagram, while $\tau$ for the original one.

In this letter we focus on the scalar effective theories in AdS with the simplest class of interaction vertices $\sum_{m=3}^\infty g_m\phi^m$, for which we present a precise general prescription for the integral recursion \eqref{eq:recursionschematic}. We comment on a preliminary application at the end.

\section{\label{sec:prescription}General Prescription}

Applying the split representation \eqref{eq:splitrepresentation} to every bulk-to-bulk propagator of a diagram, it is convenient to postpone these integrals all to the end and focus on an object $M$ alternative to $\mathcal{M}$, related by
\begin{equation}
\mathcal{M}=\int\mathcal{N}\,M,\quad
\mathcal{N}=C\prod_{i=1}^n\frac{\mathcal{C}_{\Delta_i}}{\Gamma(\Delta_i)}\prod_{a}\frac{[\mathrm{d}c_a]_{\underline\Delta_a}}{\Gamma(\pm c_a)},
\end{equation}
where $C$ is an overall constant \footnote{More precisely, $C=2^{1-2V-L}\pi^{(1-L)h}$, where $V$ and $L$ denotes the number of bulk vertices and loops, respectively.}. We call $M$ the \emph{Mellin pre-amplitude}. No $\underline\Delta$ data enter $M$ and $M\equiv M(\Delta,c,\delta)$. Correspondingly there exists an analog of \eqref{eq:recursionschematic} but for $M$.

Since we move to a different space in the loop formation, the original Mellin variables become meaningless for the new diagram and hence should all be  integrated away. However, as noted before $M$ has non-trivial dependence only on a quotient of the Mellin space parametrized by a set of Mandelstam variables. In particular, at tree level this set consists of only those associated to each bulk-to-bulk propagator (we denote them as $\Xi$). It is then equally natural to consider recursions that involve only these variables, i.e.,
\begin{equation}\label{eq:recursionguess}
M(s)=\int[\mathrm{d}\Xi]\,M_0(\Xi)\,K(\Xi,s),
\end{equation}
where $s$ denotes the Mandelstam variables appearing in the new Mellin amplitude.

Furthermore, it can be verified that $M$ has to decompose into products of contributions locally from each 1PI part of the corresponding diagram \cite{Yuan:2018qva}. Hence in principle it suffices to know the construction for arbitrary 1PI diagram, which is what we focus on in this letter.

\subsubsection{\label{sec:oneloop}One Loop}

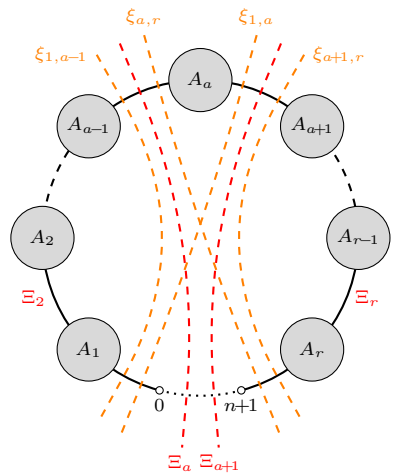
\begin{figure}[ht]
\begin{center}
\begin{tikzpicture}
\begin{scope}[scale=.7]
\draw [black,thick] (-75:3) arc [start angle=-75,end angle=0,radius=3];
\draw [black,thick,dashed] (0:3) arc [start angle=0,end angle=45,radius=3];
\draw [black,thick] (45:3) arc [start angle=45,end angle=135,radius=3];
\draw [black,thick,dashed] (135:3) arc [start angle=135,end angle=180,radius=3];
\draw [black,thick] (180:3) arc [start angle=180,end angle=255,radius=3];
\draw [black,thick,dotted] (255:3) arc [start angle=255,end angle=285,radius=3];
\draw [black,fill=black!15!white] (-45:3) circle [radius=.6];
\draw [black,fill=black!15!white] (0:3) circle [radius=.6];
\draw [black,fill=black!15!white] (45:3) circle [radius=.6];
\draw [black,fill=black!15!white] (90:3) circle [radius=.6];
\draw [black,fill=black!15!white] (135:3) circle [radius=.6];
\draw [black,fill=black!15!white] (180:3) circle [radius=.6];
\draw [black,fill=black!15!white] (225:3) circle [radius=.6];
\node [anchor=center] at (225:3) {\scriptsize $A_1$};
\node [anchor=center] at (180:3) {\scriptsize $A_2$};
\node [anchor=center] at (135:3) {\scriptsize $A_{a\!-\!1}$};
\node [anchor=center] at (90:3) {\scriptsize $A_a$};
\node [anchor=center] at (45:3) {\scriptsize $A_{a\!+\!1}$};
\node [anchor=center] at (0:3) {\scriptsize $A_{r\!-\!1}$};
\node [anchor=center] at (-45:3) {\scriptsize $A_r$};
\node [anchor=north] at (255:3) {\scriptsize $0$};
\node [anchor=north] at (285:3) {\scriptsize $n\!\!+\!\!1$};
\draw [black,fill=white] (255:3) circle [radius=2pt];
\draw [black,fill=white] (285:3) circle [radius=2pt];
\draw [red,thick,dashed] (112.5:4) -- (112.5:3) .. controls (112.5:1) and (265:1) .. (265:3) -- (265:4);
\node [anchor=center] at (265:4.3) {\scriptsize\color{red} $\Xi_a$};
\draw [red,thick,dashed] (67.5:4) -- (67.5:3) .. controls (67.5:1) and (275:1) .. (275:3) -- (275:4);
\node [anchor=center] at (275:4.3) {\scriptsize\color{red} $\Xi_{a\!+\!1}$};
\draw [orange,thick,dashed] (119.5:4) -- (119.5:3) .. controls (119.1:1) and (241:1) .. (241:3) -- (241:4);
\node [anchor=east] at (119.5:4) {\scriptsize\color{orange} $\xi_{1,a\!-\!1}$};
\draw [orange,thick,dashed] (74.5:4) -- (74.5:3) .. controls (74.5:1) and (248:1) .. (248:3) -- (248:4);
\node [anchor=south] at (74.5:4) {\scriptsize\color{orange} $\xi_{1,a}$};
\draw [orange,thick,dashed] (105.5:4) -- (105.5:3) .. controls (105.5:1) and (292:1) .. (292:3) -- (292:4);
\node [anchor=south] at (105.5:4) {\scriptsize\color{orange} $\xi_{a,r}$};
\draw [orange,thick,dashed] (60.5:4) -- (60.5:3) .. controls (60.5:1) and (299:1) .. (299:3) -- (299:4);
\node [anchor=west] at (60.5:4) {\scriptsize\color{orange} $\xi_{a\!+\!1,r}$};
\node [anchor=east] at (202.5:3) {\scriptsize\color{red} $\Xi_2$};
\node [anchor=west] at (-22.5:3) {\scriptsize\color{red} $\Xi_r$};
\end{scope}
\end{tikzpicture}
\end{center}
\vspace{-1.5em}\caption{\label{fig:polygons}Constructing a necklace diagram (the AdS boundary and all bulk-to-boundary propagagtors are omitted). Examples of channels are shown by red and orange dashed curves.}
\end{figure}

Let us first consider the construction of Mellin pre-amplitude at one loop, in which case the 1PI diagrams are the necklaces as shown in Figure \ref{fig:polygons}, where all the bulk vertices sit on the loop. We depict each vertex together with the bulk-to-boundary propagators attached to it as a grey blob, labeled by $\{A_1,\ldots,A_r\}$ in cyclic ordering. Correspondingly we denote the total conformal dimension of boundary points in each $A_a$ by $\Delta_{A_a}$. The original diagram is a chain, with two additional boundary points, $0$ and $n+1$, which are glued to form the propagator between $A_1$ and $A_r$.  The $r-1$ original bulk-to-bulk propagators associate to non-trivial OPE channels of $M_0$ (indicated by red curves), whose Mandelstam variables we denote as $\{\Xi_2,\ldots,\Xi_r\}$.

The question is then about the explicit structure of the integration kernel $K(\Xi,s)$. It turns out that if we only consider performing the $\Xi$ integrals in the recursion as exactly shown in \eqref{eq:recursionguess}, then $K$ necessarily has to be a complicated function. In order to obtain a practical kernel to work with, integrals involving extra set of original Mellin variables have to be included, which we denote as $\xi$. A natural choice for this additional set is
\begin{equation}
\{\xi_{a,b}|1\leq a\leq b\leq r\}\backslash\{\xi_{1,r}\},
\end{equation}
where each $\xi_{a,b}$ corresponds to an OPE channel that separates all the boundary points attached to $\{A_a,A_{a+1},\ldots,A_b\}$ from the rest (as illustrated by orange curves; in particular point $0$ and $n+1$ are always on the same side). It is guaranteed that $\{\Xi,\xi\}$ make up an independent set. The virtue of this choice is that, although none of these $\xi$ channels are non-trivial in the original tree, they all turn to non-trivial channels of the necklace diagram thus created. We thus label the corresponding new Mandelstam variables $s$ in the same way.

With the above setup, a practical recursion relation for the one loop construction takes the explicit form
\begin{equation}\label{eq:recursionexact}
M(s)=\int\prod_{a=2}^r\mathrm{d}\Xi_a\!\!\!\prod_{\substack{1\leq a\leq b\leq r\\(a,b)\neq(1,r)}}\!\!\!\!\mathrm{d}\xi_{a,b}\,
M_0(\Xi)\,K(\Xi,\xi;s),
\end{equation}
where $K$ naturally decomposes into $K=K_1K_2$, with
\begin{equation}\label{eq:recursionK1}
\begin{split}
K_1=&\prod_{a=1}^r\bfn{\frac{\Delta_{A_a}-\xi_{a,a}}{2}}{\frac{\Delta_{A_a}-s_{a,a}}{2}}
\prod_{a<b}^r\bfn{\frac{\xi_{a,a+1,b-1,b}}{2}}{\frac{s_{a,a+1,b-1,b}}{2}},
\end{split}
\end{equation}
with $\xi_{a,b,c,d}\equiv\xi_{a,c}+\xi_{b,d}-\xi_{a,d}-\xi_{b,c}$ and similarly for $s$, and
\begin{equation}\label{eq:recursionK2}
K_2=\prod_{a=1}^r\bfn{\frac{(\xi_{a,r}-\xi_{a+1,r})-(\Xi_a-\Xi_{a+1})}{2}}{\frac{(\xi_{a,r}-\xi_{a+1,r})-(\xi_{1,a-1}-\xi_{1,a})}{2}}.
\end{equation}
Here we used a non-standard notation for Beta function $\mathrm{B}\!\left(\substack{p\\q}\right)\equiv\frac{\Gamma(p)\Gamma(q-p)}{\Gamma(q)}$. The dependence on the $\Xi$ variables only enter $K_2$, while the new variables $s$ only enter $K_1$.

Note that depending on specific diagrams the number of integrations can vary. Whenever there is only one boundary point attached to some vertex $A_a$ we have $\xi_{a,a}=s_{a,a}=\Delta_{A_a}$, and the corresponding factor in the first product of $K_1$ is absent. Furthermore in \eqref{eq:recursionK1} and \eqref{eq:recursionK2} we always identify $\Xi_1=\Delta_0=h-c$, $\Xi_{r+1}=\Delta_{n+1}=h+c$, $\xi_{1,r}=2h$, and $\xi_{a,b}=0$ iff $a>b$.

\subsubsection{\label{sec:higherloops}Higher Loops}

While the $\xi$ variables introduced in the construction of necklace diagrams do not correspond to any non-trivial channel of the original chain diagram, they may become physical if the original diagram is some loop diagram instead. This is very encouraging as it seems to suggest that the kernel for higher loop constructions may probably follow the same pattern.

The main new ingredient needed at higher loops is a proper notion of the newly created loop. Normally this is considered as an arbitrary closed loop made up of the bulk-to-bulk propagators of the given diagram that contains the newly formed propagator. However, this is not useful for the purpose of our recursion, because a single bulk-to-bulk propagator looses its significance in the Mellin amplitude as long as it is not a tree propagator.

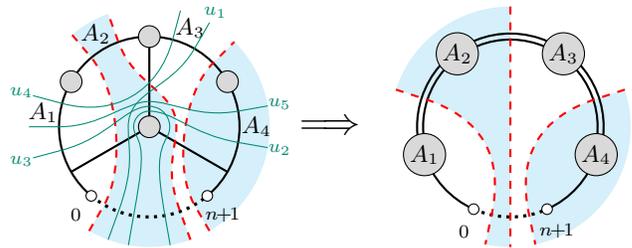
\begin{figure}[ht]
\begin{center}
\begin{tikzpicture}
\begin{scope}[scale=.8]
\begin{scope}[xshift=-3cm]
\fill [ProcessBlue!15!white] (240:2) -- (240:1.5) .. controls (240:.8) and (130:.8) .. (130:1.5) -- (130:2) arc [start angle=130,end angle=110,radius=2] -- (110:1.5) .. controls (110:.7) and (.45,.6) .. (.45,0) .. controls (.45,-.4) and (-70:.7) .. (-70:1.5) -- (-70:2) arc [start angle=-70,end angle=-120,radius=2];
\fill [ProcessBlue!15!white] (-60:2) -- (-60:1.5) .. controls (-60:.8) and (45:.8) .. (45:1.5) -- (45:2) arc [start angle=45,end angle=-60,radius=2];
\draw [black,very thick,dotted] (-130:1.5) arc [start angle=-130,end angle=-50,radius=1.5];
\draw [black,thick] (-50:1.5) arc [start angle=-50,end angle=230,radius=1.5];
\draw [black,thick] (0,0) -- (-30:1.5);
\draw [black,thick] (0,0) -- (90:1.5);
\draw [black,thick] (0,0) -- (210:1.5);
\draw [black,fill=white] (-50:1.5) circle [radius=2.5pt];
\draw [black,fill=white] (230:1.5) circle [radius=2.5pt];
\node [anchor=center] at (230:1.9) {\scriptsize $0$};
\node [anchor=center] at (-50:1.9) {\scriptsize $n\!\!+\!\!1$};
\draw [black,fill=black!15!white] (150:1.5) circle [radius=.18];
\node [anchor=center] at (172.5:1.78) {$A_1$};
\draw [black,fill=black!15!white] (0,0) circle [radius=.18];
\node [anchor=center] at (120:1.78) {$A_2$};
\draw [black,fill=black!15!white] (90:1.5) circle [radius=.18];
\node [anchor=center] at (67.5:1.78) {$A_3$};
\draw [black,fill=black!15!white] (30:1.5) circle [radius=.18];
\node [anchor=center] at (0:1.78) {$A_4$};
\draw [red,thick,dashed] (240:2) -- (240:1.5) .. controls (240:.8) and (130:.8) .. (130:1.5) -- (130:2);
\draw [red,thick,dashed] (-70:2) -- (-70:1.5) .. controls (-70:.7) and (.45,-.4) .. (.45,0) .. controls (.45,.6) and (110:.7) .. (110:1.5) -- (110:2);
\draw [red,thick,dashed] (-60:2) -- (-60:1.5) .. controls (-60:.8) and (45:.8) .. (45:1.5) -- (45:2);
\draw [PineGreen] (250:2) -- (250:1.5) .. controls (250:.7) and (-.35,-.4) .. (-.35,0) .. controls (-.35,.6) and (60:.7) .. (60:1.5) -- (60:2);
\node [anchor=center] at (60:2.2) {\scriptsize\color{PineGreen} $u_1$};
\draw [PineGreen] (260:2) -- (260:1.5) .. controls (260:.7) and ($(-150:.26)+(-60:.3)$) .. (-150:.26) arc [start angle=-150,end angle=-270,radius=.26] .. controls (.4,.26) and (-10:.7) .. (-10:1.5) -- (-10:2);
\node [anchor=center] at (-10:2.2) {\scriptsize\color{PineGreen} $u_2$};
\draw [PineGreen] (-80:2) -- (-80:1.5) .. controls (-80:.7) and ($(-30:.34)+(-120:.3)$) .. (-30:.34) arc [start angle=-30,end angle=90,radius=.34] .. controls (-.4,.34) and (195:.7) .. (195:1.5) -- (195:2);
\node [anchor=center] at (195:2.2) {\scriptsize\color{PineGreen} $u_3$};
\draw [PineGreen] (165:2) -- (165:1.5) .. controls (165:.6) and (75:.6) .. (75:1.5) -- (75:2);
\node [anchor=center] at (165:2.2) {\scriptsize\color{PineGreen} $u_4$};
\draw [PineGreen] (180:2) -- (180:1.5) .. controls (180:.8) and (-.4,.42) .. (0,.42) .. controls (.4,.42) and (10:.8) .. (10:1.5) -- (10:2);
\node [anchor=center] at (10:2.2) {\scriptsize\color{PineGreen} $u_5$};
\end{scope}
\node [anchor=center] at (0,0) {\Large $\Longrightarrow$};
\begin{scope}[xshift=3cm]
\fill [ProcessBlue!15!white] (162:2) -- (162:1.5) .. controls (162:.5) and (258:.5) .. (258:1.5) -- (258:2) arc [start angle=258,end angle=270,radius=2] -- (90:2) arc [start angle=90,end angle=162,radius=2];
\fill [ProcessBlue!15!white] (18:2) -- (18:1.5) .. controls (18:.5) and (-78:.5) .. (-78:1.5) -- (-78:2) arc [start angle=-78,end angle=18,radius=2];
\draw [black,thick] (-66:1.5) arc [start angle=-66,end angle=-18,radius=1.5];
\draw [black,thick] (-18:1.45) arc [start angle=-18,end angle=198,radius=1.45];
\draw [black,thick] (-18:1.55) arc [start angle=-18,end angle=198,radius=1.55];
\draw [black,thick] (198:1.5) arc [start angle=198,end angle=246,radius=1.5];
\draw [black,very thick,dotted] (246:1.5) arc [start angle=246,end angle=294,radius=1.5];
\draw [black,fill=black!15!white] (198:1.5) circle [radius=.35];
\node [anchor=center] at (198:1.5) {$A_1$};
\draw [black,fill=black!15!white] (126:1.5) circle [radius=.35];
\node [anchor=center] at (126:1.5) {$A_2$};
\draw [black,fill=black!15!white] (54:1.5) circle [radius=.35];
\node [anchor=center] at (54:1.5) {$A_3$};
\draw [black,fill=black!15!white] (-18:1.5) circle [radius=.35];
\node [anchor=center] at (-18:1.5) {$A_4$};
\node [anchor=center] at (246:1.9) {\scriptsize $0$};
\node [anchor=center] at (-66:1.9) {\scriptsize $n\!\!+\!\!1$};
\draw [black,fill=white] (-66:1.5) circle [radius=2.5pt];
\draw [black,fill=white] (246:1.5) circle [radius=2.5pt];
\draw [red,thick,dashed] (162:2) -- (162:1.5) .. controls (162:.5) and (258:.5) .. (258:1.5) -- (258:2);
\draw [red,thick,dashed] (90:2) -- (270:2);
\draw [red,thick,dashed] (18:2) -- (18:1.5) .. controls (18:.5) and (-78:.5) .. (-78:1.5) -- (-78:2);
\end{scope}
\end{scope}
\end{tikzpicture}
\end{center}
\vspace{-1.5em}\caption{\label{fig:channelchain}Effective diagram from a chosen chain of non-trivial channels. Double lines denote the effective tree propagator for the channels.}
\end{figure}

Instead, given that Witten diagram at any loop order admits of meaningful OPE expansion from the boundary point of view, it is natural to treat any non-trivial channel effectively as some tree bulk-to-bulk propagator. In other words, for any original diagram we can choose a maximal set of non-trivial channels which obey tree-level consistency among themselves, such that no further non-trivial channel can be included while maintaining the consistency. Then by replacing each channel by a bulk-to-bulk propagator and each region borded by them by a vertex this maximal channel set induces an effective tree diagram from the original diagram, as illustrated in Figure \ref{fig:channelchain}. Especially, for 1PI diagrams there always exists a choice such that the effective diagram is a chain, again with the two points to be glued sitting at the ends.

Once we obtain an effective chain diagram, we are able to introduce its corresponding $\Xi$ and $\xi$ variables following exactly what we did at one loop. The recursion formula is then the same as \eqref{eq:recursionexact}.

Take Figure \ref{fig:channelchain} as an example, the independent integration variables under consideration are thus
\begin{equation}\label{eq:intvariables2loop}
\begin{split}
\{&\Xi_2,\Xi_3,\Xi_4,\\
&\xi_{1,1},{\color{magenta}\xi_{1,2}},{\color{magenta}\xi_{1,3}},\xi_{2,2},{\color{magenta}\xi_{2,3}},{\color{magenta}\xi_{2,4}},\xi_{3,3},\xi_{3,4},\xi_{4,4}\}.
\end{split}
\end{equation}
With these variables we read off the kernel $K$ from the effective diagram using the same formulas \eqref{eq:recursionK1} and \eqref{eq:recursionK2}. In this specific case, except for $\{\xi_{1,2},\xi_{1,3},\xi_{2,3},\xi_{2,4}\}$ all the other five $\xi$ variables correspond to non-trivial OPE channels of the original two loop diagram.

It may at first sight cause some worry that the original pre-amplitude $M_0$ depends on extra Mandelstam variables not belonging to \eqref{eq:intvariables2loop}, which are $\{u_1,u_2,\ldots,u_5\}$, as indicated by green curves in Figure \ref{fig:channelchain} (the reason that all these variables enter $M_0$ will be discussed in detail in \cite{Yuan:2018qva}). However, all these variables turn out to linearly depend on the ones listed in \eqref{eq:intvariables2loop}. Explicitly,
\begin{align}
u_1=&\;\xi_{2,2}-\xi_{2,3}+\xi_{3,3}+\Xi_2-\Xi_3+\Xi_4,\\
u_2=&\;\Delta_{n+1}+\xi_{2,2}-\xi_{2,4}+\xi_{3,4}+\Xi_2-\Xi_3,\\
u_3=&\;\Delta_0+\xi_{1,1}-\xi_{1,2}+\xi_{2,2}-\Xi_2+\Xi_3,\\
u_4=&\;\xi_{1,1}-\xi_{1,2}+\xi_{1,3}+\xi_{2,2}-\xi_{2,3}+\xi_{3,3},\\
u_5=&\;2h+\xi_{1,1}-\xi_{1,2}+\xi_{2,2}-\xi_{2,4}+\xi_{3,4}.
\end{align}
Hence in the recursion formula \eqref{eq:recursionexact} there are no extra $u$ integrations apart from those of $\Xi$ and $\xi$ in \eqref{eq:intvariables2loop}, and correspondingly in the original pre-amplitude $M_0(\Xi,\xi,u)$ we substitute the $u$ variables using the above relations.

This linear dependence among the Mandelstam variables for different non-trivial OPE channels is a general feature that occurs in the construction of higher-loop diagrams. This is in fact consistent with the existence of different maximal chain of non-trivial channels, each of which induces some effective diagram that can be equally used for the construction.

\begin{figure}[ht]
\begin{center}
\begin{tikzpicture}
\begin{scope}[scale=.7]
\draw [black,very thick,dotted] (-60:1.5) arc [start angle=-60,end angle=-120,radius=1.5];
\draw [black,thick] (-60:1.5) arc [start angle=-60,end angle=240,radius=1.5];
\draw [black,thick] (180:1.5) -- (180:2.5);
\node [anchor=center] at (180:2.7) {$4$};
\coordinate (p1) at ($(90:1.5)+(120:1)$);
\node [anchor=center] at ($(0,0)!1.1!(p1)$) {$1$};
\coordinate (p2) at ($(90:1.5)+(60:1)$);
\node [anchor=center] at ($(0,0)!1.1!(p2)$) {$2$};
\draw [black,thick] (p1) -- (90:1.5) -- (p2);
\draw [black,thick] (0:1.5) -- (0:2.5);
\node [anchor=center] at (0:2.7) {$3$};
\node [anchor=center] at (240:1.9) {\scriptsize $h\!-\!c_3$};
\node [anchor=center] at (-60:1.9) {\scriptsize $h\!+\!c_3$};
\node [anchor=center] at (146:1.75) {\scriptsize $c_1$};
\node [anchor=center] at (34:1.75) {\scriptsize $c_2$};
\fill [black] (180:1.5) circle [radius=2.5pt];
\fill [black] (90:1.5) circle [radius=2.5pt];
\fill [black] (0:1.5) circle [radius=2.5pt];
\draw [black,fill=white] (-60:1.5) circle [radius=2.5pt];
\draw [black,fill=white] (-120:1.5) circle [radius=2.5pt];
\draw [red,thick,dashed] (157.5:2.5) -- (157.5:1.5) .. controls (157.5:.5) and (-100:.5) .. (-100:1.5) -- (-100:2);
\node [anchor=center] at ($(0,0)!1.15!(157.5:2.5)$) {\scriptsize\color{red}$t_1$};
\draw [red,thick,dashed] (22.5:2.5) -- (22.5:1.5) .. controls (22.5:.5) and (-80:.5) .. (-80:1.5) -- (-80:2);
\node [anchor=center] at ($(0,0)!1.15!(22.5:2.5)$) {\scriptsize\color{red}$t_2$};
\draw [orange,thick,dashed] (135:2.5) -- (135:1.5) .. controls (135:.5) and (-30:.5) .. (-30:1.5) -- (-30:2.5);
\node [anchor=center] at ($(0,0)!1.2!(-30:2.5)$) {\scriptsize\color{orange}$t_5|\Delta_4$};
\draw [orange,thick,dashed] (45:2.5) -- (45:1.5) .. controls (45:.5) and (210:.5) .. (210:1.5) -- (210:2.5);
\node [anchor=center] at ($(0,0)!1.2!(210:2.5)$) {\scriptsize\color{orange}$t_3|\Delta_3$};
\draw [orange,thick,dashed] (112.5:2.5) -- (112.5:1.5) .. controls (112.5:.5) and (67.5:.5) .. (67.5:1.5) -- (67.5:2.5);
\node [anchor=center] at ($(0,0)!1.15!(67.5:2.5)$) {\scriptsize\color{orange}$t_4|S$};
\end{scope}
\end{tikzpicture}
\end{center}
\vspace{-1.5em}\caption{\label{fig:triangle4pt}Constructing the 4-point triangle diagram.}
\end{figure}
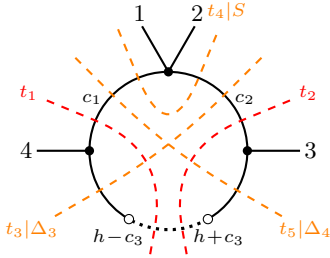

\section{\label{sec:application}Applications}

We now apply our recursive construction to a simple example, the 4-point triangle diagram with a quartic vertex (When $h=1$ or $h=2$, with a single scalar of $\Delta=2$, explicit results were worked out in \cite{Aharony:2016dwx}). We choose to construct this diagram according to Figure \ref{fig:triangle4pt}. Associated to the orange curves variables before the slash refer to the original $\xi$ variables while those after for the corresponding new $s$ variables. Hence in applying \eqref{eq:recursionK1} and \eqref{eq:recursionK2} we have $r=3$, together with the substitution
\begin{equation}
\begin{split}
&\Xi_1=h-c_3,\quad
\Xi_2=t_1,\quad
\Xi_3=t_2,\quad
\Xi_4=h+c_3\\
&\Delta_{A_1}=\Delta_4,\quad
\Delta_{A_2}=\Delta_{12},\quad
\Delta_{A_3}=\Delta_3,\\
&\xi_{1,1}=s_{1,1}=s_{2,3}=\Delta_4,\quad
\xi_{1,2}=t_3,\quad
\xi_{2,2}=t_4,\\
&\xi_{2,3}=t_5,\quad
\xi_{3,3}=s_{3,3}=s_{1,2}=\Delta_3,\quad
s_{2,2}=S,
\end{split}
\end{equation}
where $\Delta_{12}\equiv\Delta_1+\Delta_2$. 
It can also be verified that the original tree has the Mellin pre-amplitude
\begin{widetext}
\begin{equation}\label{eq:M0triangle4pt}
\begin{split}
M_0=&\frac{\Gamma(\frac{\Delta_4-c_3\pm c_1}{2})}{\Gamma(\frac{\Delta_4+h-c_3-t_1}{2})}\,\Gamma({\textstyle\frac{h\pm c_1-t_1}{2}})\,\Gamma({\textstyle\frac{\Delta_{12}\pm c_1\pm c_2}{2}})\,
\Gamma({\textstyle\frac{h\pm c_2-t_2}{2}})\,\frac{\Gamma(\frac{\Delta_3+c_3\pm c_2}{2})}{\Gamma(\frac{\Delta_3+h+c_3-t_2}{2})}\,F,
\end{split}
\end{equation}
with $F$ being a function free of poles, defined by another Mellin integral
\begin{equation}
F=\int\frac{\mathrm{d}t_6}{2\pi i}\frac{\Gamma(\frac{h\pm c_1-t_6}{2})\Gamma(\frac{h-c_2+t_1-t_2-t_6}{2})\Gamma(\frac{t_6-t_1}{2})\Gamma(\frac{\Delta_{12}-h+c_2+t_6}{2})}{\Gamma(\frac{h\pm c_1-t_1}{2})\Gamma(\frac{h-c_2-t_2}{2})\Gamma(\frac{\Delta_{12}\pm c_1+c_2}{2})\Gamma(\frac{\Delta_{12}+t_1-t_2}{2})\Gamma(\frac{\Delta_{12}+h-c_2-t_6}{2})}.
\end{equation}
\end{widetext}
This representation is consistent with existing tree Feynman rules \cite{Fitzpatrick:2011ia,Paulos:2011ie} via contour deformation. So altogether we have a $6$-fold $t$ integrals for this diagram.

It is in principle not hard to perform numerical evaluations following \cite{Czakon:2005rk,Hahn:2004fe}. Furthermore, the fact that these are Mellin integrals allows for convenient contour analysis, which concludes that $M$ is
\begin{equation}
\begin{split}
&\Gamma({\textstyle\frac{\Delta_3\pm c_2\pm c_3}{2}})\Gamma({\textstyle\frac{\Delta_4\pm c_3\pm c_1}{2}})\Gamma({\textstyle\frac{\Delta_{12}\pm c_1\pm c_2}{2}})\Gamma({\textstyle\frac{\Delta_{1234}}{2}\!-\!h})\\
&\times\Gamma({\textstyle\frac{2h-\Delta_3\pm c_2\pm c_3}{2}})\Gamma({\textstyle\frac{2h-\Delta_4\pm c_3\pm c_1}{2}})\Gamma({\textstyle\frac{2h\pm c_1\pm c_2-S}{2}})
\end{split}
\end{equation}
times another function that is again free of poles. In other words, the entire pole structure of $M$ is encoded in the Gamma functions above. Combining with the remaining spectrum integrals, it is easy to observe that the Mellin amplitude $\mathcal{M}$ contains only one non-trivial OPE channel, corresponding to poles at
\begin{equation}
S=\underline\Delta_{1}+\underline\Delta_{2}+2k,\quad k=0,1,\ldots,
\end{equation}
which verifies the presence of the double-trace operators $[\underline{\mathcal{O}}_1\underline{\mathcal{O}}_2]_{k,\ell}$ in the $S$-channel OPE with twist at the corresponding location \cite{Aharony:2016dwx}. 

It is also convenient to extract expressions for the residues. There are four contributions to the residue at the physical poles, from each $\Gamma(\frac{2h\pm c_1\pm c_2-S}{2})$ with a specific choice of the two signs. It turns out that they all yields the same result and that the corresponding residue contour localizes all the $t$ integrals as well as the $c_1$ and $c_2$ integrals. As a result, for instance, the residue at the leading pole $S=\underline\Delta_{12}$ is
\begin{widetext}
\begin{equation}\label{eq:triangleleadingresidue}
\begin{split}
\residue{S=\underline\Delta_{12}}\mathcal{M}=&
-8\pi^{2h}\prod_{i=1}^4\frac{\mathcal{C}_{\Delta_i}}{\Gamma(\Delta_i)}\prod_{a=1}^2\frac{\mathcal{C}_{\underline\Delta_a}}{\Gamma(\underline\Delta_a)}
\times\underbrace{\Gamma({\textstyle\frac{\Delta_{12}+\underline\Delta_{12}}{2}-h})}_{\parbox{2.5cm}{\tikz{
\begin{scope}[scale=.4]
\draw [black,very thick] (0,0) circle [radius=1.5];
\draw [black,thick] (120:1.5) -- (0,0) -- (-120:1.5);
\begin{scope}[xshift=.75cm]
\draw [black,thick] (0:.75) .. controls (60:.6) and (120:.6) .. (180:.75) .. controls (-120:.6) and (-60:.6) .. (0:.75);
\end{scope}
\fill [black] (0,0) circle [radius=2pt];
\node [anchor=center] at (120:2) {\tiny $\mathcal{O}_2$};
\node [anchor=center] at (-120:2) {\tiny $\mathcal{O}_1$};
\node [anchor=center] at (0:3.1) {\tiny $[\underline{\mathcal{O}}_{1}\underline{\mathcal{O}}_{2}]_{0,0}$};
\end{scope}
}}}\times
\int\frac{[\mathrm{d}c_3]_{\underline\Delta_3}}{\Gamma(\pm c_3)}\underbrace{\frac{\Gamma(\frac{\underline\Delta_1\pm(h-\Delta_4)\pm c_3}{2})\Gamma(\frac{\underline\Delta_2\pm(h-\Delta_3)\pm c_3}{2})}{\Gamma(h+\frac{\underline\Delta_{12}-\Delta_{34}}{2})\Gamma(\frac{\underline\Delta_{12}\pm(\Delta_3-\Delta_4)}{2})}}_{\parbox{2.5cm}{\tikz{
\begin{scope}[scale=.4]
\draw [black,very thick] (0,0) circle [radius=1.5];
\draw [black,thick] (180:1.5) -- (90:.7) -- (60:1.5);
\draw [black,thick] (180:1.5) -- (-90:.7) -- (-60:1.5);
\draw [black,thick] (-90:.7) -- (90:.7);
\fill [black] (90:.7) circle [radius=2pt];
\fill [black] (-90:.7) circle [radius=2pt];
\node [anchor=center] at (180:3.1) {\tiny $[\underline{\mathcal{O}}_{1}\underline{\mathcal{O}}_{2}]_{0,0}$};
\node [anchor=center] at (60:2) {\tiny $\mathcal{O}_3$};
\node [anchor=center] at (-60:2) {\tiny $\mathcal{O}_4$};
\node [anchor=west] at (0,0) {\tiny $c_3$};
\end{scope}
}}}.
\end{split}
\end{equation}
\end{widetext}
In the above, the two under-braces highlighted two expressions, each of which turns out to be the pre-amplitude of a 3-point diagram involving the double-trace primary $[\underline{\mathcal{O}}_1\underline{\mathcal{O}}_2]_{0,0}$ on the boundary, as indicated below the braces. These can easily be verified by straightforward computation. Hence we explicitly observed a factorized structure of the leading pole residue.

\section{\label{sec:discussion}Discussion}

In this letter we presented recursion relations among scalar Witten diagrams with arbitrary $\phi^m$ interaction vertices, which help to construct Mellin integral representation for Mellin amplitudes at arbitrary loops that can be directly read off from the diagram following general formulas \eqref{eq:recursionK1} and \eqref{eq:recursionK2}. To illustrate its practical convenience we also provided one application in the $4$-point triangle diagram with a quartic vertex, where the complete pole structure were derived and residue at the leading physical pole explicitly worked out. Detailed derivations and discussions as well as further applications are to be discussed in a forthcoming companion paper \cite{Yuan:2018qva}. The resulting representation from our method in general involves multidimensional Mellin integrals, and some improvements to the usual contour analysis are in fact needed in order for an efficient analysis of its analytic properties, which are also to be introduced in \cite{Yuan:2018qva}. Here it is worth to point out that in all our discussions we have implicitly assumed that $n\leq d+2$, so that the dimension of the Mellin space stays as $\frac{n(n-3)}{2}$ and no extra subtlety occurs to the definition of $\mathcal{M}$ in \eqref{eq:MellinDef} (see, e.g., \cite{Penedones:2016voo}). It would be interesting to further explore the $n>d+2$ case.

\begin{acknowledgments}
\emph{Acknowledgments.} EYY would like to thank Nima Afkhami-Jeddi, Nima Arkani-Hamed, Carlos Cardona, Bartek Czech, Song He, Johannes Henn, Yu-tin Huang, Juan Maldacena, Mauricio Romo, Shu-heng Shao, Marcus Spradlin, Anastasia Volovich, and in particular Shota Komatsu and Eric Perlmutter, for useful discussions. 
EYY is supported by the US Department of Energy under grant DE-SC0009988 and by a Carl P.~Feinberg Founders Circle Membership.
\end{acknowledgments}




\bibliography{wdiagrams}

\end{document}